 \documentstyle[12pt]{article}
\def\a{\alpha}
\def\b{\beta}
\def\g{\gamma}
\def\d{\delta}
\def\e{\epsilon}

\def\l{\lambda}
\def\L{\Lambda}
\def\m{\mu}
\def\n{\nu}
\def\r{\rho}

\def\P{\Pi}
\def\s{\sigma}
\def\S{\Sigma}

\def\O{\Omega}
\def\L{\Lambda}
\catcode`\@=11
\newfam\msbfam
\font\twelvemsb=msbm10 scaled\magstep1
\font\tenmsb=msbm10 \font\sevenmsb=msbm7 
\textfont\msbfam=\twelvemsb
\scriptfont\msbfam=\tenmsb \scriptscriptfont\msbfam=\sevenmsb
\def\Bbb{\relax\ifmmode\expandafter\Bbb@\else
        \expandafter\nonmatherr@\expandafter\Bbb\fi}
\def\Bbb@#1{{\Bbb@@{#1}}}
\def\Bbb@@#1{\fam\msbfam\relax#1}
\catcode`\@=\active

\begin{document}

\title{ About a Super-Ashtekar-Renteln Ansatz}
\author{G\'eza F\"ul\"op\\ Institute of Theoretical Physics\\
S-412 96 G\"oteborg\\ Sweden}
\date{September 1993}

\maketitle

\vskip -10cm
\noindent \hskip 10cm G\"oteborg, ITP 93-7
\vskip 2mm
\noindent \hskip 10cm   May,1993
\vskip 2mm
\noindent \hskip 10cm   Revised September,1993
\vskip +10cm

\begin{abstract} The Ashtekar-Renteln Ansatz gives the self-dual solutions to
the Einstein equation. A direct generalization of the Ashtekar-Renteln
An\-satz to N=1 supergravity is given both in the canonical and in
the covariant formulation and a geometrical property of the solutions
is pointed out.
 \end{abstract}

\section{Introduction}
   The Ashtekar-Renteln Ansatz is a simple solution to all the
constraints of
gravitation in the Hamiltonian formulation using Ashtekar's variables
for
the case of non-vanishing cosmological constant \cite{ash}.
 This Ansatz
gives the solutions of the Einstein equation corresponding to a
vanishing Weyl
tensor. The Hamiltonian formulation of gravity using Ashtekar's
variables uses the following constraint system: The Hamilton constraint
that "pushes" the spacelike hypersurface forward in time:
\begin{equation}
{\cal H} = \frac{i}{2} f^{ij}_k E^a_i E^b_j F^k_{ab} +
\frac{\lambda}6  f^{ijk} {\epsilon}_{abc}  E^a_i  E^b_j  E^c_k \approx 0
\end{equation}
where $ F^i_{ab}$ is the field strength of the configuration space
variable $A^i_a$; $E^a_i$ is the canonically conjugated momentum
and $\lambda$ is the cosmological constant. a,b,c... are the space
indices; i,j,k... are the internal indices. The vector constraints that
generate the spacelike diffeomorphisms look like:
\begin{equation}
{\cal H}_a = E^b_i F^i_{ab} \approx 0
\end{equation}
 Finally the constraints called Gauss' law are:
\begin{equation}
{\cal G}_i = D_a E^a_i \approx 0
\end{equation}
The Ashtekar-Renteln Ansatz gives a simple relation between the `electric'
field $E^a_i$ and the `magnetic' field $B^a_i$ obtained from the field
strength $F^i_{ab}$; a relation that solves all the constraints:
\begin{equation}
E^a_i = - \frac{3i}{\lambda} B^a_i
\end{equation}
As we see it is crucial to have a non-vanishing cosmological constant.
The field configuration obtained by using the Ashtekar-Renteln Ansatz
has self-dual field strength and it contains some geometrical information
as it will become clear later.

   There is a relation equivalent to the  Ashtekar-Renteln Ansatz
related to the CDJ-action \cite{cdj}. This pure connection action
is built up from tensors of the form:
\begin{equation}
{\Omega}_{ij} = {\epsilon}^{\a \b \g \d} F_{\a \b i} F_{\g \d j}
\label{omega}
\end{equation}
$\a$, $\b$... being space-time indices. The Ashtekar-Renteln Ansatz
is equivalent to demanding that:
\begin{equation}
\O_{ij} \propto \d_{ij} \label{odelta}
\end{equation}
The equivalence can be seen from the definition of the conjugated
momenta for $A^i_a$.

   The goal of the present
paper is to generalize these results to  $N=1$ supergravity which was
formulated in terms of Ashtekar's variables in \cite{jacobson}.
For some more results see also \cite{nicolai},\cite{sano1} and
\cite{sano2}.The structure of this paper is the following: in
Section 2 after introducing the
notation a generalization of the Ashtekar-Renteln Ansatz is given.
Then a generalization of the $\Omega$ matrix is obtained together with
some conditions (generalizations of eq.(\ref{odelta})), which when applied
to the generalized $\Omega$, lead to the generalized Ashtekar-Renteln
Ansatz. This Ansatz will be called the super-Ashtekar-Renteln Ansatz.
Afterwards a geometrical interpretation of these
results is briefly described, and it is shown that the field strength
obtained by the super-Ashtekar-Renteln Ansatz (or equivalently by
applying the conditions generalized from eq.(\ref{odelta}) on the generalized
 $\Omega$) is self-dual and for non-degenerate metric it forms a basis
for all self-dual two-forms on the manifold. In Section 3 the covariant
form of the generalized Ansatz is given together with the covariant form
of the above mentioned conditions. Section 4 contains some general comments
about the self-duality of the solutions and a wave function of the
Chern-Simons type discussed in \cite{kodama},\cite{sano1} and \cite{sano2}.
The similarity between the super and the non-super
case is stressed all along the paper.

\section{ The phase space form of the
\newline ``super-Ashtekar-Renteln Ansatz"}
   The phase space action of supergravity is:
\begin{equation}
S= \int \dot{A}_a^i E^a_i -\dot{\Psi}_{aA} \P^{aA} - {\Lambda}^i {\cal G}_i
- N {\cal H} - N^a {\cal H}_a - \n_{A} S^{\dag A} -\m_A S^A
\end{equation}
where the basic variables are the ``old" $A_a^i$ and $E^a_i$ and the
Grassmann odd
vectorspinor $\Psi_{aA}$ with its conjugate momenta $\P^{aA}$.
One can of course also begin with a Hilbert-Palatini type Lagrangian using
the variables $\psi_{aA}$ and their complex conjugate
$\bar{\psi}_{aA'}$. One can then arrive at the
momenta $\P^{aA}$ as in \cite{jacobson}. The point is that one can compute
$\bar{\psi}_{aA'}$ knowing $\P^{aA}$ and $\n_{A}$.

   The conventions
used for the Pauli matrices are the following:
\begin{equation}
[ \s_i, \s_j ]_{-} = \sqrt{2} f_{ijk} \s^k
\end{equation}
\begin{equation}
[ \s_i, \s_j ]_{+~B}^{~A} = \e^A_{~B} \d_{ij}
\end{equation}

In the case when the spinor indices are not explicitly written the
following convention is used:
\[ \Psi_{a} \s_i \P^a  \equiv \Psi_{aA} \s_{i~B}^{~A} \P^{aB} \]

 The constraints are these:

-Gauss' law:
\begin{equation}
{\cal G}_i = D_a E^a_i + \frac{i}{\sqrt{2}} \Psi_{a} \s_i \P^a \approx 0
\label{sgauss}
\end{equation}
-Another constraint is the one usually called the left supersymmetry
generator:
 \begin{equation}
S = D_a \P^{a}  -i \a E^a_i \s_{i} \Psi_{a} \label{susy.l}
\end{equation}
-The constraint which in some representation is the complex conjugate of
$S^A$  is $S^{\dag A}$ and called the right supersymmetry generator:
\begin{equation}
S^{\dag } = f_{ijk}  E^a_i  E^b_j \s_k [ -2 \sqrt{2} D_a \psi_b
-i \bar{\a} \e_{abc} \P^c]
\end{equation}
-The Hamilton and the vector constraints have the same role of generators
of diffeomorphisms as in the non-super case. They look like:
\begin{equation}
{\cal H}_a = E^b_i F^i_{ab} - 2 \P^b D_{[a} \psi_{b]} -
i \a \psi_a \s_i E^b_i \psi_b
 \approx 0
 \end{equation}
and
\[
{\cal H} = \frac{i}{2} \e_{abc} f^{ijk} E^a_i E^b_j B^c_k -
\frac12  \a \bar{\a} f^{ijk} \e_{abc} E^a_i E^b_j E^c_k
+ i 2 \sqrt{2} \P^b E^a_i \s^i D_{[a} \psi_{b]} \]
\begin{equation}
+ \bar{\a} \e_{abc} \P^a E^b_i \s^i \P^c\\
- \frac{1}{2} \a f^{ijk} E^a_i E^b_j \psi_a \s_k \psi_b \approx 0
\end{equation}
Two remarks about
this set of constraints might be useful. The first one is that
this form of the constraints is equivalent to the form in \cite{jacobson},
though the notation is different. This notation was used
in \cite{bengtsson}. The second remark concerns the constants
$\a$ and its complex conjugate ${\bar \a}$ which give the
cosmological constant: $\l = - \a \bar{\a}$.
The first generalization of super gravity to the case of non-vanishing
cosmological constant was given in \cite{townsend}. It turned out that
the cosmological constant had to be negative. This means that it can be
written as minus the square of a constant, see \cite{jacobson}.
Since there is no evident reason why this constant ($\a$) should be real,
it seems reasonable to use a more general complex constant.
The reality condition then implies that the cosmological constant
should have the above mentioned form. The non-super case can be
recovered by setting all the Grassmann odd variables equal to zero.

   The algebra between Gauss' law and the left supersymmetry generator looks
like:
\begin{equation}
\{ {\cal G}_i, {\cal G}_j \} = i f_{ij}^{~~k} {\cal G}_k
\end{equation}
\begin{equation}
\{ {\cal G}_i, S^A \} = - \frac{i}{\sqrt{2}} \s_{i~B}^{~A} S^B
\end{equation}
\begin{equation}
\{ S^A, S^B \} = -i \a \s_i^{~AB} {\cal G}_i
\end{equation}
This is evidently a semisimple graded algebra (GSU(2)). For a treatment of
these algebras see e.g. \cite{pais}.  Therefore the configuration space is
spanned by a graded-SU(2) connection and it is natural to introduce a notation
that reflects this fact.
 The ``super" configuration
space variables can be defined as:
\begin{equation}
{\bf A}_a^{\bar{\imath}} = ( A_a^i, \psi^A_a)
\end{equation}
\begin{equation}
{\bf E}^a_{\bar{\imath}} = ( E^a_i, \P^a_A)
\end{equation}
The barred indices are the supersymmetry indices $ \bar{\imath} =  (i,A)$.
The field strength for ${\bf A_a^{\bar{\imath}}}$ is:
\begin{equation}
{\bf F}_{ab}^{\bar{\imath}} = ({\Bbb F}_{ab}^i, 2 D_{[a} \psi_{b]}^A)
\label{superfield}
\end{equation}
where
\begin{equation}
{\Bbb F}_{ab}^i =  F_{ab}^i + i \a \psi_a \s^i \psi_b \label{fieldstrength}
\end{equation}
This field strength ${\Bbb F}_{ab}^i$ becomes the ``usual"
field strength when the cosmolo\-gi\-cal constant vanishes. Using it one can
write the constraints in a somewhat simpler form.

One can now introduce a ``super"-covariant derivative and a ``super"-Gauss'
law:
\begin{equation}
{\bf D}_a {\bf E}^a_{\bar{\imath}} = \partial_a {\bf E}^a_{\bar{\imath}} +
\L_{\bar{\imath} \bar{\jmath}}^{~~\bar{k}} {\bf A}_a^{\bar{\jmath}}
 {\bf E}^a_{\bar{k}} \approx 0 \label{sugauss}
\end{equation}
where:
\[ \L_{ij}^{~~k} = i f_{ij}^{~~k} \]
\[ \L_{iA}^{~~B} = \frac{i}{\sqrt{2}} \s_{iA}^{~~B} \]
\[ \L_{AB}^i = -i \a \s^i_{AB} \]
\[ \L_{Ai}^{~~B} = - \frac{i}{\sqrt{2}} \s_{iA}^{~~B} \]
This constraint is of course Gauss' law for ${\bar{\imath}} = i,j,k$
and the left
supersymmetry generator for ${\bar{\imath}} = A,B$. Another notation that
will be used is:
\begin{equation}
{\bf B}^a_{\bar{\imath}} = ({\Bbb B}^{a}_i, \b^a_A)
\end{equation}
where:
\begin{equation} {\Bbb B}^{a}_i = \frac12 \e^{abc} {\Bbb F}_{bc}^i \label{magn}
\end{equation}
\begin{equation} \b^a_A = \e^{abc}  D_{[b} \psi_{c]A} \end{equation}
(Note that ${\Bbb F}$ used here is the field strength defined by eq.
 (\ref{fieldstrength}).) The Hamilton and the vector constraints now take the
form:
\begin{equation}
{\cal H} = \frac{i}{2} \e_{abc} f^{ijk} E^a_i E^b_j ({\Bbb B}^c_k +
i \a {\bar \a} E^c_k)
+ \e_{abc} E^a_i \P^b \s^i (i {\sqrt 2} \b^c - {\bar \a} \P^c) \approx 0
\end{equation}

\begin{equation}
{\cal H}_a = E^b_i {\Bbb F}^i_{ab} - 2 \P^b  D_{[a} \psi_{b]} \approx 0
\end{equation}
    In order to solve the constraints one would like to find the
``super-electric" field as a function of the ``super-magnetic" field.
One can  start by trying to find the solutions of the following form:
\begin{equation}
{\bf E}^a_{\bar{\imath}} = M^{\bar{\imath \jmath}}
{\bf B}^a_{\bar{\jmath}}
\label{e}
\end{equation}
where $M^{\bar{\imath \jmath}} $ is a $5x5$ matrix. Since
${\Bbb B}^{a}_i$ and $\b^a_A$
are independent of each other the vector constraints impose some
conditions on the
matrix $M$. Namely that the $3x3$ submatrix $M^{ij}$ is symmetric,
the $2x2$ submatrix
is antisymmetric and the ``mixed" $2x3$ and $3x2$ parts are just the
transposed of each
other. Writing now equation (\ref{e}) in the original notation:
\begin{equation}
E^{ai} = M^{(ij)} {\Bbb B}^{a}_j + M^{iA} \b^a_A
\end{equation}
\begin{equation}
\P^{aA} = M^{Ai} {\Bbb B}^{a}_i +\frac12 \e^{AB} M_D^{~D} \b^a_B
\end{equation}
\newpage
A simple Ansatz one can make for this matrix is that the first $3x3$
submatrix is
diagonal and the ``mixed" part vanishes. One obtains then:
\begin{equation}
E^a_i =  \frac{i}{\a \bar{\a}} {\Bbb B}^{a}_i \label{sara1}
\end{equation}
\begin{equation}
\P^{aA} = \frac{ i \sqrt{2}}{\bar{\a}} \e^{abc} D_{[b} \psi_{c]}^A
\label{sara2}
\end{equation}
${\Bbb B}$ is defined here as in eq.(\ref{magn}).
This is our generalization of the Ashtekar-Renteln Ansatz and it
will be referred to as the ``super-Ashtekar-Renteln" Ansatz. This Ansatz
solves
trivially all the constraints as it can be seen using eq.(\ref{superfield}).
One can express the time-space components
of the field strength from the equations of motion by varying the phase space
action by the `electric' field:
\begin{equation}
{\bf F}_{0a \bar{\imath}} = N \{{\bf A}_{a \bar{\imath}}, {\cal H} \}
+ N^b \{{\bf A}_{a \bar{\imath}}, {\cal H}_b \} +
\{{\bf A}_{a \bar{\imath}},\n_A S^{\dag A} \} \label{newfield}
\end{equation}

   This could be done in the non-super case as well. There one could insert
the result obtained from the equation of motion for $F_{0ai}$ in the matrix
$\O$ defined in equation (\ref{omega}). If one inserts here the form of the
`electric' field from
the Ashtekar-Renteln Ansatz one obtains equation (\ref{odelta}). Or vice-versa:
demanding that $\O_{ij} \propto \d_{ij}$ one can obtain a relation
between $E$ and the field strength. This relation was exactly the
Ashtekar-Renteln Ansatz.

   A similar structure can be found in supergravity too.
 One can define a matrix
${\bf \O}$  as:
\begin{equation}
{\bf \O}_{\bar{\imath} \bar{\jmath}} = \e^{\a \b \g \d}
{\bf F}_{\a \b \bar{\imath}}
 {\bf F}_{\a \b \bar{\jmath}}
\end{equation}

Inserting the super-Ashtekar-Renteln Ansatz here one can compute $N$:
\begin{equation}
N =  \frac{\a \bar{\a}}{24 \mid {\Bbb B} \mid} \O_{ii} \label{N}
\end{equation}
$N^a$ can be computed from another matrix. Define ${\bf W}$ as:
\begin{equation}
{\bf W}_{\bar{\imath} \bar{\jmath}} = {\bf F}_{0a \bar{\imath}}
{\bf B}^a_{\bar{\jmath}}
\end{equation}
${\bf \O}$ is in fact 8 times the symmetric part of ${\bf W}$. $N^a$ from
here:
\begin{equation}
N^a = \frac12 W_{[ij]} \e^{abc} ({\Bbb B}^{-1})^i_c ({\Bbb B}^{-1})^j_b
\label{Na}
\end{equation}

To complete the solution one has to find $\n_A$. This is most easily done
from
the
``mixed" part of ${\bf \O}$ (${\bf \O}_{iA}$) by changing its tangent space
index i into
spinor indices.  One can always find a two-spinor
corresponding to a space-time vector:
\begin{equation}
v_{AA'} = v_I \s^I_{AA'}
\end{equation}
and the Lorentz symmetry of the space-time vector space transforms into the
SL(2,$\bf{C}$)
symmetry of the spinor space. In our case however one only has space
indices
and the
rotation symmetry SO(3). This means we only need SU(2) spinors and the
two-spinor
corresponding to a space vector is:
\begin{equation}
v_{AB} = v_i \s^i_{AB}
\end{equation}
where the $\s^i$-s are the Pauli matrices. In exactly the same way one can
define:
\begin{equation}
{\bf \O}_{DEA} = {\bf \O}_{iA} \s^i_{DE}
\end{equation}
This ${\bf \O}_{DEA}$ can be decomposed in irreducible parts. Since it is
symmetric in the last two indices the decomposition looks like:
\begin{equation}
{\bf \O}_{DEA} = {\bf \O}_{(DEA)} + \frac12 {\bf \O}_{D[EA]} +
\frac12 {\bf \O}_{E[DA]}
\end{equation}
The totally symmetric part vanishes when using the super-Ashtekar-Renteln
Ansatz and the decomposition becomes:
\begin{equation}
{\bf \O}_{DEA} = \frac13 \e_{AD} {\bf \O}^B_{~EB} +
\frac13 \e_{AE} {\bf \O}^B_{~DB}
\end{equation}
and one obtains:
\begin{equation}
\n_A = \frac{i \a^2 \bar{\a}}{4 \mid {\cal B} \mid} {\bf \O}^B_{~AB}
\end{equation}
So the super-Ashtekar-Renteln Ansatz gives us a set of solutions and some
constraints
on the ${\bf W}$ matrix: its symmetric tracefree part vanishes as well as
the totally
symmetric part of the ${\bf \O} = {\bf \O}_i \s^i$ tensor.
\newpage
   Similarly to the non-super case one could start in searching for a
solution by demanding that:
\begin{equation}
  {\hat{\bf \O}}_{ij} = 0, \label{eq1}
\end{equation}
 where ${\hat{\bf \O}}$ is the traceless part of ${\bf \O}$

\begin{equation}
   {\bf \O}_{AB} \propto \e_{AB} \label{eq2}
\end{equation}
\begin{equation}
  {\bf \O}_{i(A} \s^i_{BC)} =0 \label{eq3}
\end{equation}

\hskip -6mm and solving the equations of motion one obtains for the
electric field
and the spinor momentum the same form as given by the
super-Ashtekar-Renteln Ansatz given above. These three relations can then
be understood as a covariant formulation of the super-Ashtekar-Renteln
Ansatz.

     There is a geometrical interpretation of these results related to the
interpretation of the results of the Ashtekar-Renteln Ansatz in the
non-super
case \cite{torre}. There exists a one to one correspondence between
the way
the duality operator acts on two-forms and the conformal structure on
a four dimensional space. Knowing the metric up to a conformal factor is
enough to
compute how the duality operator acts and vice-versa. This fact is
expressed
in four dimensions by Urbantke's formula \cite{urbantke}, \cite{harnett}
which gives the conformal structure as a function of a basis of the
self-dual
vector fields. If the field strength is self-dual and it is non-degenerate
it can be used as a basis to all self-dual fields, that is
it gives the metric up to a conformal factor by Urbantke's formula:
\begin{equation}
 g^{\m \n} \propto f_{ijk} \e^{\m \a \b \g} \e^{\d \n \r \s}  F_{\a \b}^i
 F_{\g \d}^j F_{\r \s}^k \label{u}
\end{equation}
The field configuration obtained using the Ashtekar-Renteln Ansatz has
self-dual
field strength. This can be seen by writing the self-dual part of the
curvature
tensor as a sum of its irreducible components and inserting the solution
of
the Einstein equation. The field strength is non-degenerate if the
determinant
of the ``magnetic" field does not vanish, as is the case for generic
solutions. This means we can use
equation (\ref{u}). The conformal factor can be determined from the action.

   One can use similar arguments to understand the nature of solutions in
supergravity.  Since it is not
clear (at least to the author) how to decompose the curvature tensor in
this case, it seems now easier to go the other way around and try instead
to use
Urbantke's formula and see what kind of a result it gives. The main
question is:
does Urbantke's formula give a tensor that has the structure of the
metric in
the ADM decomposition? The ADM decomposition expressed in Ashtekar's
variables looks like:
\begin{equation}
(-g)^{\frac12} g^{\a \b} = \left( \begin{array}{cc} -N^{-1} & N^{-1} N^b \\
N^{-1} N^a & N E^a_i E^b_i - N^{-1} N^a N^b \end{array} \right)
\label{decomp}
\end{equation}
Inserting the form of ${\Bbb F}_{0ai}$ from the equations of motion
(\ref{newfield}) into Urbantke's formula (\ref{u}) the obtained tensor in
general is not of the form eq.(\ref{decomp}). The time-time and the
time-space components of the matrix depend
explicitly on the `electric' field. Using the super-Ashtekar-Renteln
Ansatz to express E in terms of the field strength ${\Bbb F}$ one can
again
compute the time-time and the time-space components and notice that they
are
of the same form as in equation (\ref{decomp}). This means that Urbantke's
formula
gives the conformal structure for those solutions of the constraints
that obey the super-Ashtekar-Renteln Ansatz.
{}From here one can con\-clude  that the su\-per-Ash\-te\-kar-Ren\-teln
Ansatz
gives a set of self-dual solutions to the constraint system
of N=1 supergravity and when the `magnetic' field is non-degenerate,
that is corresponding to a non-degenerate metric, this set of field
strengths is a basis for all self-dual vector fields.

   To summarize: in contrast to the non-super case the field strength
 ${\Bbb F}$ does not in general correspond to a self-dual solution,
but the field strengths obtained through the super-Ashtekar Ansatz do
correspond to self-dual solutions.

\section{The covariant form}

   The Ashtekar-Renteln Ansatz (in the non-super case) can be formulated
in a covariant form too \cite{samuel}, \cite{cdjm}. To derive this form
one can start from an action:

\begin{equation}
I = \int \frac12 {\S}^{AB} \wedge F_{AB} - \l {\S}^{AB} \wedge {\S}_{AB}
- {\Psi}_{(ABCD)} {\S}^{AB} \wedge {\S}^{CD}
\end{equation}
The fundamental variables are the field-strength two-forms and
the two-forms $\S$ which are obtained as:
\begin{equation}
{\S}_{\a \b}^{AB} = e_{[\a}^{AA'} e_{\b]A'}^{~B} \label{action1}
\end{equation}
The last term in the action with the totally symmetric Lagrange
multiplier ${\Psi}_{(ABCD)}$ is to ensure that $\S$ is of the form of eq.
(\ref{action1}). $\l$ is the cosmological constant. The following Ansatz
is the covariant formulation of the Ashtekar-Renteln Ansatz:
\begin{equation}
\frac12 F^{AB}_{\a \b} = \l {\S}^{AB}_{\a \b} \label{cov.ns}
\end{equation}

  A very similar construction can be made in the super case and one
can obtain a generalization of eq.(\ref{cov.ns}). This generalization will
be the covariant form of the super-Ashtekar-Renteln Ansatz.

   The action one can start with is the one given in \cite{jacobson}
and \cite{cdjm}:
\begin{displaymath}
I = \int  [ \frac{i}2 \S^{AB} \wedge F_{AB}
-i \sqrt{2} {\chi}^A \wedge D {\psi}_{A}
- \frac12 {\Psi}_{(ABCD)} {\S}^{AB} \wedge \S^{CD}
- {\O}_{(ABC)} \S^{AB} \wedge {\chi}^C
\end{displaymath}
\begin{equation}
+ \frac{\a \bar{\a}}{2 \sqrt2} \S^{AB} \wedge \S_{ AB}
+ \frac{\a}2 \S^{AB} \wedge {\psi}_{A} \wedge {\psi}_{B}
+ \frac{\bar{\a}}{2} {\chi}^A \wedge {\chi}_{A}]
\end{equation}

The new variable here is:
\begin{equation}
{\chi}^A_{\a \b} = e_{[ \a}^{AB'} {\bar{\psi}}_{\b] B'} \label{chi}
\end{equation}
This variable is related to the momenta of the ${\psi}_{\a A}$ in the
canonical formulation. The term with the totally symmetric Lagrange-
multiplier ${\O}_{(ABC)}$ is there to ensure that eq.(\ref{chi}) holds.
Details about
how ${\Psi}_{(ABCD)}$ and ${\O}_{(ABC)}$ work can be found in \cite{cdjm}.
$\a$ and $\bar{\a}$ give the cosmological constant as before.
It is straight forward to show that this action is invariant under
the left super-symmetry transformation (eq.(\ref{susy.l})) if one writes the
variations in the following form:
\begin{equation}
{\d}_{\e} \S^{AB} = i {\chi}^{(A} {\e}^{B)}
\end{equation}
\begin{equation}
{\d}_{\e} {\psi}^A = D {\e}^A
\end{equation}
\begin{equation}
{\d}_{\e} A^{AB} =  i \a {\psi}^{(A} {\e}^{B)}
\end{equation}
 \begin{equation}
{\d}_{\e} {\chi}^A = \frac{i \a}{\sqrt2} {\S}^{AB} {\e}_B
\end{equation}
(The space-time indices are suppressed.) One can now obtain the equations
of motion by varying the action with respect to the different variables.
If one varies first with respect to $A$ one obtains the
equation corresponding to the Gauss' law (eq. (\ref{sgauss})).
Variation with respect to $\psi$ gives the left super-symmetry
generator (eq. (\ref{susy.l})). In one sentence: variation with respect
to the super configuration-space variable ${\bf A}_a^{\bar{\imath}}$
gives us the "super"-Gauss' law (eq. (\ref{sugauss})).
The variation with respect to $\S$ gives the following equation:
\begin{equation}
F_{\a \b AB} + 2 i {\Psi}_{(ABCD)} {\S}_{\a \b}^{CD}
+ 2 i {\O}_{(ABC)} {\chi}^C_{\a \b}
- i \sqrt2 \a \bar{\a} {\S}_{\a \b AB}
- i \a {\psi}_{[\a A} {\psi}_{\b] B} = 0 \label{eqm1}
\end{equation}
Varying by $\chi$ one obtains:
\begin{equation}
- i \sqrt2 D_{[\a} {\psi}_{\b] A} - {\O}_{(ABC)} {\S}_{\a \b}^{BC}
+ \bar{\a} {\chi}_{\a \b A} = 0 \label{eqm2}
\end{equation}
Introducing a new notation in the first equation of motion (eq. (\ref{eqm1}))
\[{\Bbb F}_{\a \b}^{AB} = F_{\a \b}^{AB}
- i \a {\psi}_{[\a}^{A} {\psi}_{\b]}^{B} \]
one can notice the following: the Ansatz
\begin{equation}
{\Bbb F}_{\a \b}^{AB} = i \sqrt2 \a \bar{\a} {\S}_{\a \b AB} \label{sara1.cov}
\end{equation}
and
\begin{equation}
{\chi}_{\a \b A} = \frac{i \sqrt2}{\bar{\a}} D_{[\a} {\psi}_{\b] A}
\label{sara2.cov}
\end{equation}
is equivalent to demanding that:
\begin{equation}
{\Psi}_{(ABCD)} = 0 \label{sara1.eq}
\end{equation}
and
\begin{equation}
{\O}_{(ABC)} = 0 \label{sara2.eq}
\end{equation}
Equations (\ref{sara1.cov}) and (\ref{sara2.cov}) are the covariant formulation
of the super-Ashtekar-Renteln Ansatz. In a more elegant way:
\[ {\Bbb F} = i \sqrt2 \a \bar{\a} {\S}\]
and \[ {\chi} = \frac{i \sqrt2}{\bar{\a}} D \wedge {\psi} \]
The other two equations (eq. (\ref{sara1.eq}) and (\ref{sara2.eq})) are
equivalent to equations (\ref{eq1}), (\ref{eq2}) and (\ref{eq3}).
The easiest way to see that this covariant formulation of the
super-Ashtekar-Renteln Ansatz is equivalent to the phase-space
formulation expressed in equations (\ref{sara1}) and (\ref{sara2}) is to
use the gauge where $e_a^0 = 0$.

\section{Final comments}

   When quantizing a theory one wants to obtain wave function(s)
that satisfy the constraint operators corresponding the
classical constraints. In the connection representation
of gravity one needs wave functions depending on the connection:
$\Psi[A]$. The variables become operators acting on the wave
function: $ {\hat A}^i_a \Psi[A] =  A^i_a \Psi[A] $ and
$ {\hat E}^a_i \Psi[A] = \frac{\d}{\d A^i_a} \Psi[A] $.
There are two unsolved problems that one meets here:
regularization and factor ordering. For a discussion about
the possible choices of factor ordering and wave functions
corresponding to them see \cite{brugman}. One can choose
the ordering so that the functional derivatives are at the left
and the functions of the connection at the right. This is
called factor ordering I in \cite{brugman}. There is a relatively
simple wave function that solves the Gauss' law and the
Hamilton constraint in the non-super case. It is the
exponential of the Chern-Simons term divided by the cosmological
constant. This means that here too it is crucial to have a
non-vanishing cosmological constant.

In super gravity we have of course two more fundamental variables
that appear as operators in the quantum theory:
$\hat{\psi}_{aA} \Psi = {\psi}_{aA} \Psi$ and
$\hat{\P}^{aA} \Psi = \frac{\d}{\d {\psi}_{aA}} \Psi$.
Now a wave function very similar to the one existing in the non-super
theory solves quite trivially the Gauss' law, the left and right
super symmetry and the Hamilton constraints if the super-Ashtekar-Renteln
Ansatz holds \cite{kodama},\cite{sano1}, \cite{sano2}. This is true if the
factor ordering
is chosen as "factor ordering 1" in \cite{brugman}. The vector constraint
is however not eliminated by this wave function. Since in this ordering
the vector constraint operator does not generate space-like diffeomorphisms
this does not affect the diffeomorphism invariance of the wave function.
This wave function has the following explicit form:
\begin{equation}
S = e^{ - \frac{i}{2 \a \bar{\a}} \int {\e}^{abc} Tr [A_a {\partial}_b A_c
+ \frac{\sqrt2 i}{3} A_a A_b A_c + i \a A_a \psi_b \psi_c
+\sqrt2 \a \psi_a \partial_b \psi_c]}
\end{equation}

   As we see there are many aspects of supergravity related to the
super-Ashtekar-Renteln Ansatz which are simple generalizations
of the non-super case: the form of the Ansatz is quite similar to
the non-super Ansatz both in the canonical and the covariant
formulation; with a proper choice of the fundamental variables
we just have a GSU(2) constraint algebra instead of SU(2);
one can also obtain almost similar conditions put on tensors
depending on the `magnetic' field, conditions that are equivalent
to the Ansatz and one can obtain wave functions  that are in
principle similar to each other. (By similar it is meant that
the super case is a super-generalization of the non-super one.)
What is different however is the self-duality of the different
solutions. In pure gravity the field strength was self-dual by
construction while in super-gravity the super-Ashtekar-Renteln
Ansatz had to be imposed in order to obtain self-dual solutions
and this was also
reflected in the geometrical interpretation of the Ansatzes.
   The presence of spinor variables in the super case leads also
to a non-vanishing torsion. This can be computed but its form is
quite complicated.

 \paragraph{ACKNOWLEDGEMENTS}

 I am grateful to In\-ge\-mar Bengts\-son and an anonymous referee
for ma\-ny use\-ful sugges\-tions and for improvements in the manuscript.
I want also to thank Peter Peld\'an for interesting discussions.

\end{document}